\documentclass[aps,prl,twocolumn,groupedaddress,amsmath,amssymb]{revtex4}
\usepackage{graphicx,color}

\newcommand{\ET}{\mbox{$\not \hspace{-0.10cm} E_T$ }}

\begin{document}


\title{Dark sector shining through 750 GeV dark Higgs boson at the LHC}


\author{P. Ko}
\email[]{pko@kias.re.kr}
\affiliation{School of Physics, KIAS, Seoul 02455, Korea}

\author{Takaaki Nomura}
\email[]{nomura@kias.re.kr}
\affiliation{School of Physics, KIAS, Seoul 02455, Korea}


\date{\today}

\begin{abstract}
We consider a dark sector with $SU(3)_C \times U(1)_Y \times U(1)_X$ and three families of 
dark fermions that are chiral under dark $U(1)_X$ gauge symmetry, whereas scalar dark matter 
$X$ is the SM singlet.  $U(1)_X$ dark symmetry is spontaneously broken by nonzero VEV of 
dark Higgs field $\langle \Phi \rangle$, generating the masses of dark fermions and dark photon $Z^\prime$. 
The resulting dark Higgs boson $\phi$ can be produced at the LHC by dark quark loop 
(involving 3 generations) and will decay into a pair of photon through  charged dark fermion loop. 
Its decay width can be easily $\sim 45$ GeV due to its possible decays into a pair of dark 
photon, which is not strongly constrained by the current LHC searches $pp \rightarrow \phi \rightarrow 
Z^\prime Z^\prime$ followed by $Z^\prime$ decays into the SM fermion pairs. 
The scalar DM can achieve thermal relic density without  conflict with direct detection 
bound or the invisible $\phi$ decay into a pair of DM.
\end{abstract}


\maketitle

\section{Introduction}
%
Recently both ATLAS and CMS Collaborations announced that there are some excess  in the 
diphoton channel around $m_{\gamma\gamma} \approx 750$ GeV
~\cite{ATLAS-CONF-2015-081,CMS:2015dxe}:
\begin{eqnarray}
\sigma ( pp \rightarrow \phi \rightarrow \gamma\gamma ) & = & (6.2^{+2.4}_{-2.0})\ {\rm fb} \quad  
({\rm ATLAS})
\\
& = & (5.6\pm{2.4})\ {\rm fb}  \quad ({\rm CMS} )
\\
\Gamma_{\rm tot} ( \phi ) & \sim &  45{\rm GeV} ({\rm ATLAS}) 
\end{eqnarray}
whereas the CMS data prefers  a smaller decay width~\cite{CMS:2015dxe}.
Furthermore, at Moriond 2016, ATLAS and CMS have reported that the local (global) 
significances of the diphoton excess are about 3.9(2.0)$\sigma$ and 3.4(1.6)$\sigma$, 
respectively,   where CMS added 0.6 fb$^{-1}$ new data to the 13 TeV analysis and combined 
with 8 TeV  data~\cite{ATLAS, CMS:2016owr}. 

This excess motivated a lot of phenomenological study on possible scenarii of new physics beyond the Standard Model (BSM) which include models related to DM physics~\cite{Mambrini:2015wyu, Backovic:2015fnp,Franceschini:2015kwy,Ellis:2015oso,Dutta:2015wqh,No:2015bsn,Chao:2015ttq,Bian:2015kjt,Bai:2015nbs,Chao:2015nsm,Han:2015cty,Bi:2015uqd,Cho:2015nxy,Cline:2015msi,Bauer:2015boy,Barducci:2015gtd,Dey:2015bur,Dev:2015isx,Moretti:2015pbj,Patel:2015ulo,Davoudiasl:2015cuo,Han:2015yjk,Park:2015ysf,Li:2015jwd,Chao:2015nac,Chiang:2015tqz,Huang:2015svl,Kanemura:2015bli,Hernandez:2015hrt,Kaneta:2015qpf,Nomura:2016fzs,Ko:2016lai,Ghorbani:2016jdq,Palle:2015vch,Modak:2016ung,Deppisch:2016scs,Berlin:2016hqw,Bhattacharya:2016lyg,D'Eramo:2016mgv, Borah:2016uoi}, 
new gauge symmetry models~\cite{Chao:2015nsm,Dey:2015bur,Patel:2015ulo,Ko:2016lai,Modak:2016ung,Deppisch:2016scs,Berlin:2016hqw,Borah:2016uoi, Martinez:2015kmn,Chang:2015bzc,Feng:2015wil,Boucenna:2015pav,Hernandez:2015ywg,Pelaggi:2015knk,deBlas:2015hlv,Huang:2015rkj,Das:2015enc,Cao:2015scs,Jiang:2015oms,Dasgupta:2015pbr,Karozas:2016hcp}
and other models~\cite{Harigaya:2015ezk,Angelescu:2015uiz,Nakai:2015ptz,Knapen:2015dap,Buttazzo:2015txu,Pilaftsis:2015ycr,DiChiara:2015vdm,Higaki:2015jag,McDermott:2015sck,Low:2015qep,Bellazzini:2015nxw,Gupta:2015zzs,Petersson:2015mkr,Molinaro:2015cwg,Cao:2015pto,Matsuzaki:2015che,Kobakhidze:2015ldh,Cox:2015ckc,Becirevic:2015fmu,Demidov:2015zqn,Fichet:2015vvy,Curtin:2015jcv,Chakrabortty:2015hff,Ahmed:2015uqt,Agrawal:2015dbf,Csaki:2015vek,Falkowski:2015swt,Aloni:2015mxa,Gabrielli:2015dhk,Benbrik:2015fyz,Kim:2015ron,Alves:2015jgx,Megias:2015ory,Carpenter:2015ucu,Bernon:2015abk,Arun:2015ubr,Chakraborty:2015jvs,Ding:2015rxx,Han:2015dlp,Han:2015qqj,Luo:2015yio,Chang:2015sdy,Bardhan:2015hcr,Antipin:2015kgh,Wang:2015kuj,Cao:2015twy,Huang:2015evq,Liao:2015tow,Heckman:2015kqk,Dhuria:2015ufo,Kim:2015ksf,Berthier:2015vbb,Chala:2015cev,Murphy:2015kag,Belyaev:2015hgo,Badziak:2015zez,Chakraborty:2015gyj,Cao:2015xjz,Altmannshofer:2015xfo,Cvetic:2015vit,Gu:2015lxj,Allanach:2015ixl,Craig:2015lra,Cheung:2015cug,Liu:2015yec,Zhang:2015uuo,Casas:2015blx,Hall:2015xds,Salvio:2015jgu,Chway:2015lzg,Son:2015vfl,Tang:2015eko,An:2015cgp,Cao:2015apa,Wang:2015omi,Cai:2015hzc,Kim:2015xyn,Gao:2015igz,Bi:2015lcf,Goertz:2015nkp,Anchordoqui:2015jxc,Dev:2015vjd,Bizot:2015qqo,Ibanez:2015uok,Kang:2015roj,Hamada:2015skp,Kanemura:2015vcb,Low:2015qho,Marzola:2015xbh,Ma:2015xmf,Jung:2015etr,Potter:2016psi,Palti:2016kew,Han:2016bus,Danielsson:2016nyy,Chao:2016mtn,Csaki:2016raa,Hernandez:2016rbi,Dutta:2016jqn,Ito:2016zkz,Zhang:2016pip, Sahin:2016lda,Fichet:2016pvq,Stolarski:2016dpa}.
It is not easy to generate a large enough width $\sim 45$ GeV 
with large $BR ( \phi \rightarrow \gamma\gamma)$, maintaining relevant cross section of $\sigma ( p p \rightarrow \phi
\rightarrow \gamma\gamma ) \sim O(10)$ fb and evading various collider search bounds.  

In this letter, we solve these problems by introducing dark U(1)$_X$ gauge symmetry, 
dark photon $Z^\prime$,   three generations of dark fermions with $SU(3)_C \times U(1)_Y$ 
charges and singlet scalar DM $X$. 
Dark photon $Z'$ can decay into SM fermions via a small $Z-Z'$ mixing.  
Dark fermions are assumed to be chiral under U(1)$_X$ dark gauge symmetry and get massive 
after spontaneous breaking  of  U(1)$_X$ by nonzero VEV of U(1)$_X$-charged complex scalar field $\Phi$, and a new Higgs boson $\phi$ appears from $\Phi$.  This simple setup 
for dark matter is a viable DM scenario with interesting signatures at high energy colliders.

%
%
\section{Model}
%
\begin{center} 
\begin{table}[b]
\begin{tabular}{|c||c|c|c|c|c|c|c|c||c|c|}\hline\hline  
&\multicolumn{8}{c||}{Fermions} & \multicolumn{2}{c|}{Scalar} \\\hline
& ~$E_L$~ & ~$E_R^{}$~ & ~$N^{}_{L}$ ~ & ~$N_R$~  & ~$U_L$~ & ~$U_R$  & ~$D_L$  & ~$D_R$ & ~$\Phi$~ & ~$X$~ \\\hline 
SU(3) & $\bf{1}$ & $\bf{1}$  & $\bf{1}$ & $\bf{1}$ & $\bf{3}$ & $\bf{3}$  & $\bf{3}$   & $\bf{3}$ & $\bf{1}$   & $\bf{1}$ \\\hline 
SU(2) & $\bf{1}$ & $\bf{1}$  & $\bf{1}$ & $\bf{1}$ & $\bf{1}$ & $\bf{1}$  & $\bf{1}$   & $\bf{1}$ & $\bf{1}$   & $\bf{1}$ \\\hline 
U(1)$_Y$ & $-1$ & $-1$  & $0$ & $0$ & $\frac{2}{3}$ & $\frac{2}{3}$ & $\frac{-1}{3}$  & $\frac{-1}{3}$ & $0$ & $0$  \\\hline
U(1)$_X$ & $a$ & $-b$  & $-a$ & $b$ & $-a$ & $b$ & $a$  & $-b$ & $a+b$ & $a$  \\\hline
\end{tabular}
\caption{Contents of new fermions and scalar fields and their charge assignments under the gauge 
symmetry SU(3)$\times$SU(2)$_L\times$U(1)$_Y\times$U(1)$_X$.  
 We consider three families of dark fermions.}
\label{tab:1}
\end{table}
\end{center}
Let us  introduce a dark sector with new dark fermions which carry both the SM 
$SU(3)_C \times U(1)_Y$  quantum numbers as well as dark U(1)$_X$ gauge charges, 
and a SM singlet complex scalar field $X$ as summarized in Table~\ref{tab:1}.  
In this model, every right-handed fermion $f_R$ in the SM has its partner 
fermion $F_L$ with nonzero dark charge in the dark sector. Then the $\overline{F}_L f_R$ 
operator becomes invariant under the SM gauge transformation. Its nonzero dark charge is 
cancelled by the dark charge of scalar 
DM $X$ in such a way that $\overline{F}_L f_R X$ becomes gauge invariant operator.
And $F_L$ becomes vectorlike under the SM gauge group by introducing its chiral partner $F_R$. 
The model is very simple and free from gauge anomalies for arbitrary $a$ and $b$.  
A novel feature of this model is that the new fermions $F_L$ and $F_R$ are {\it chiral} under dark 
U(1)$_X$ gauge symmetry so that they are massless before spontaneous symmetry breaking. 
And their effects on $\phi \rightarrow gg, \gamma\gamma$ through triangle diagram evades from 
the decoupling  theorem as their mass becomes heavy.

The Yukawa interactions and the scalar potential including new fields in the dark sector are described by 
\begin{align}
L_{\rm Yukawa} =& y^E \bar E_L E_R \Phi + y^N \bar N_L N_R \Phi^\dagger 
+ y^U \bar U_L U_R \Phi^\dagger \nonumber \\
&+ y^D \bar D_L D_R \Phi  + y^{Ee} \bar E_L e_R X + y^{Uu} \bar U_L u_R X^\dagger \nonumber \\
& + y^{Dd} \bar D_L d_R X + h.c., 
\end{align}
\begin{align}
V =& \mu^2 H^\dagger H + \lambda (H^\dagger H)^2 + \mu_\Phi^2 \Phi^\dagger \Phi 
+ \mu_X^2 X^\dagger  X \nonumber \\
&+ \lambda_\Phi (\Phi^\dagger \Phi)^2+ \lambda_X (X^\dagger X)^2   + \lambda_{H\Phi} 
(H^\dagger H)(\Phi^\dagger \Phi) \nonumber \\
& + \lambda_{HX} (H^\dagger H)(X^\dagger X)+ \lambda_{X\Phi} (X^\dagger X)(\Phi^\dagger \Phi).
\end{align}
where $H$ denote the SM Higgs field \footnote{
For $a=b=1$, there appears an extra term $\Phi^\dagger X^2$ in the potential, 
which breaks U(1)$_X$  down to $Z_2$ subgroup after $S$ develops nonzero VEV.  
Likewise, for $3a = (a+b)$, there appears an extra term $\Phi^\dagger X^3$, which breaks 
U(1)$_X$  down to $Z_3$ subgroup after $S$ develops nonzero VEV.  
In this paper, we do not consider these possibilities,    relegating the readers to 
Ref.~\cite{Baek:2014kna}  and Refs.~\cite{Ko:2014nha,Ko:2014loa,Guo:2015lxa} 
for $Z_2$ and $Z_3$ cases, respectively.}.  
We have suppressed the generation indices on the SM and the dark fermions for simplicity. 
The Yukawa interactions provide mass terms for the dark fermions $F$, which decay  
through $F \to X f$.   
$X$ is the SM singlet and can be a good DM candidate. 
Note that there is an accidental $Z_2$ symmetry, $X\rightarrow -X$, $F_L \rightarrow 
- F_L$ and $F_R \rightarrow - F_R$ which make $X$ stable at renormalizable level. 
There could be gauge invariant operators that break this accidental $Z_2$ symmetry: 
$X^\dagger \Phi^n$ and/or $X \Phi^n$ which would generate nonzero VEV for $X$ after U(1)$_X$ symmetry 
breaking by nonzero $\langle \Phi \rangle \neq 0$.  Gauge invariance requires that 
$\pm a/(a+b) = n$ to be an integer.  We can forbid this type of operators by making a judicious 
choice of $a,b$ so that $\pm a/(a+b)$ is not an integer. Or we can make $n$ very large so that 
even if $X$ develops a nonzero VEV, the lifetime of $X$ becomes long enough 
($\tau_X \gtrsim 10^{28}$ sec) to be a good DM candidate. 
This model can be considered as a generalization of the singlet portal 
extensions of the SM where dark matter lives in the dark sector~\cite{Baek:2013qwa}, 
but the dark sector now contains dark fields which are charged under the SM gauge group 
as well as dark gauge group,   unlike the earlier models~\cite{Baek:2013qwa}.  

The gauge symmetry is broken after $H$ and $\Phi$ get non-zero VEVs:
\begin{equation}
H= \left(\begin{array}{cc}
 G^+    \\
\frac{1}{\sqrt{2}} ( v+ h + iG^0)     
\end{array}
  \right)\,,  \quad
\Phi = \frac{1}{\sqrt{2}} (v_\phi + \phi + i G_\phi),
\end{equation}
where $G^\pm$, $G^0$ and $G_\phi$ are NG bosons which are absorbed by $W^\pm$, 
$Z$ and $Z'$ respectively.  We shall call $\phi$ as dark Higgs boson, since it appears 
as a result of spontaneous breaking of dark U(1)$_X$ gauge symmetry. 

We assume $\lambda_{H\Phi}$ is negligible and the mixing between SM Higgs boson 
$h$ and $\phi$ is negligibly small which is consistent with the current Higgs data analysis
~\cite{Chpoi:2013wga}.    Then the scalar VEVs are given approximately by 
\begin{equation}
v \simeq \sqrt{\frac{-\mu^2}{\lambda}}, \quad v_\Phi \simeq \sqrt{\frac{-\mu_\Phi^2}{\lambda_\Phi}}.
\end{equation}
The masses of new fermions are generated such that
\begin{equation}
M_F = \frac{y^F}{\sqrt{2}} v_\Phi \ ,
\end{equation}
where $F=E,N,U$ and $D$.

We consider kinetic mixing of the U(1)$_Y$ and U(1)$_X$ gauge fields which are denoted respectively as $\tilde B_\mu$ and $\tilde X_\mu$; 
\begin{align}
\mathcal{L}_{\text{kin}} =& -\frac{1}{4} W^a_{\mu \nu} W^{a \mu \nu} \nonumber \\
&- \frac{1}{4}(\tilde{B}_{\mu\nu},\tilde{X}_{\mu\nu})
\left(
\begin{array}{cc}
1 & s_\chi\\
s_\chi & 1
\end{array}
\right)
\left(
\begin{array}{c}
\tilde{B}^{\mu\nu}\\
\tilde{Z}^{'\mu\nu}
\end{array}\right), \label{kin} 
\end{align}
where $s_\chi \equiv \sin \chi$.
 The kinetic terms are diagonalized by the following non-unitary transformation;
 \begin{align}
\left(
\begin{array}{c}
\tilde{B}_\mu\\
\tilde{X}_\mu
\end{array}\right)=
\left(
\begin{array}{cc}
1 & -t_\chi\\
0 & 1/t_\chi
\end{array}\right)
\left(
\begin{array}{c}
B_\mu\\
X_\mu
\end{array}\right) ~,
\end{align}
where $t_\chi = \tan \chi$.
After $\Phi$ and $H$ develop non-zero VEVs, the mass matrix for neutral gauge field is approximately given by
\begin{equation}
\frac{1}{8} \left( \begin{array}{c} \tilde{Z} \\ X \end{array} \right)^T 
\left( \begin{array}{cc}  (g^2 +g'^2) v^2 & t_\chi g' \sqrt{g^2 + g'^2} v^2 \\ t_\chi g' \sqrt{g^2 + g'^2} v^2 &  4(a+b)^2 g_X^2 v_\Phi^2 \end{array} \right)
 \left( \begin{array}{c} \tilde{Z} \\ X \end{array} \right).
\end{equation}
where $W^3_\mu = \cos \theta_W Z_\mu + \sin \theta_W A_\mu$ and $B_\mu = -\sin \theta_W + \cos \theta_W A_\mu$ are used.
Assuming $\chi \ll 1$ \footnote{The upper bound on the kinetic mixing is roughly $\lesssim 0.01$ in the dark 
photon mass range $m_{Z^\prime} \lesssim 350$ GeV considered in this letter~\cite{Jaeckel:2012yz}.}, 
neutral gauge boson masses are  
\begin{equation}
m_Z^2 \simeq \frac{1}{4}(g^2+g'^2) v^2, \quad m_{Z'}^2 \simeq (a+b)^2 g_X^2 v_\Phi^2.
\end{equation}
The mass eigenstates are given by
\begin{align}
\left( \begin{array}{c} Z_\mu\\ Z'_\mu
\end{array}\right)=
\left(
\begin{array}{cc}
\cos \theta & -\sin \theta \\
\sin \theta & \cos \theta \end{array} \right)
\left( \begin{array}{c}
\tilde{Z}_\mu \\ X_\mu \end{array}\right) ~,
\end{align}
and the small $Z-Z'$ mixing angle is given by
\begin{equation}
\tan 2 \theta \simeq \frac{g' \sqrt{g^2 + g'^2} v^2}{2 (m_Z^2 - m_{Z'}^2 ) } t_\chi.
\end{equation}
\begin{figure}[tbh] 
\begin{center}
\includegraphics[width=70mm]{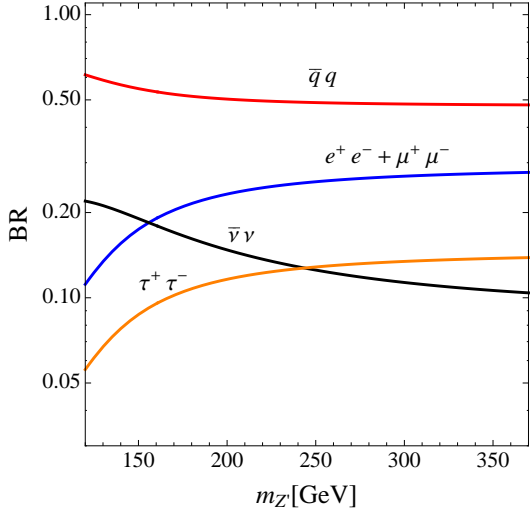} \hspace{5mm}
\caption{Branching ratios of $Z^\prime$ as a function of $m_{Z^\prime}$. }
\label{fig:BRZp}
\end{center}
\end{figure}
In Fig.~\ref{fig:BRZp}, we show the branching ratios of $Z^\prime$ as a function of its mass. 
Here $q=u,d,s,c,b$,   and $\nu\bar{\nu}$ includes all the three flavors. Note that $Z^\prime$ 
decays into the SM through the  kinetic mixing so that $\Gamma (Z^{'} )/m_{Z^\prime} 
\sim O(\chi^2) \lesssim 10^{-4}$. Therefore $Z^\prime$ would be a very narrow resonance.

We also find that Yukawa coupling of new dark fermions and $\lambda_\Phi$ can be written 
in terms of $g_X$ and $m_{Z'}$; 
\begin{align}
\label{eq:yF}
y^F &= \frac{\sqrt{2} (a+b) g_X M_F}{m_{Z}'}, \\
\label{eq:lamS}
\lambda_\Phi &= \frac{(a+b)^2 m_\phi^2 g_X^2}{2 m_{Z'}^2}.
\end{align}
In our analysis, we require these couplings are perturbative as $y^F < 4 \pi$ and 
$\lambda_\Phi < 4 \pi$.

\section{Phenomenology}
\subsection{750 GeV Diphoton Excess}
\begin{figure}[tbh] 
\begin{center}
\includegraphics[width=70mm]{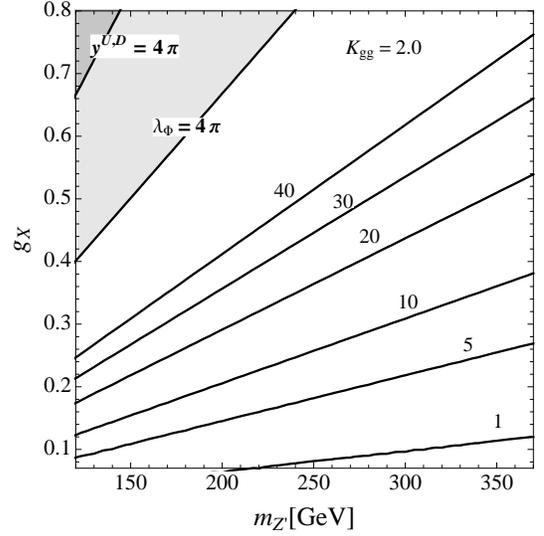} \hspace{5mm}
\caption{ The $\sigma (gg \to \phi)$ in unit of pb where 3 copy of fermions in Table~\ref{tab:1} 
are applied and $a \simeq b \simeq 1$ with $a\neq b$ is adopted. 
We used parameter set as $\{M_{U,D}, M_{E,N}, m_X, \lambda_{X\Phi} \} = \{800 \, {\rm GeV}, 
400 \, {\rm GeV}, 350 \, {\rm GeV}, 0.075 \}$.
Gray(light gray) region indicate $y^{U,D}(\lambda_{\Phi}) > 4 \pi$ using Eq.~(\ref{eq:yF}) and (\ref{eq:lamS}).}
\label{fig:XS}
\end{center}
\end{figure}

%
In this section, we analyze the production of $\phi$ and its decays at the LHC 13 TeV.
The production of $\phi$ is through gluon fusion process where the relevant effective coupling 
is given by
  \begin{equation}
{\cal L}_{\phi gg} = \frac{\alpha_s}{8\pi } \left( \sum_{F=U,D} \frac{(a+b) \sqrt{2} g_X }{m_{Z'}} A_{1/2}(\tau_F)  \right) 
\phi G^{a\mu \nu}G^a_{\mu \nu} \,,\label{eq:LggS}
 \end{equation}
 where $A_{1/2}(\tau) =  2 \tau [1+(1-\tau) f(\tau)]$ with $f(\tau) = [\sin^{-1} (1/\sqrt{\tau})]$ 
 for $\tau \geq 1$  and  $\tau_F \equiv 4 m_F^2/m_\phi^2$.
 We find that the effective coupling is described by $m_{Z'}$ and $g_X$ since exotic fermion 
 mass is given by VEV of $\Phi$.
 Applying the effective coupling, the production cross section for the dark Higgs $\phi$
  is calculated by use of 
 CalcHEP~\cite{Belyaev:2012qa} with {\tt CTEQ6L} PDF~\cite{Nadolsky:2008zw}.
 Fig.~\ref{fig:XS} shows the cross section in the $m_{Z'}-g_X$ plane using parameter setting  
 $\{M_{U,D}, M_{E,N}, m_X, \lambda_{X\Phi} \} = \{800 \, {\rm GeV}, 400 \, {\rm GeV}, 350 \, 
 {\rm GeV}, 0.075 \}$ 
 as a reference and K-factor for gluon fusion as $K_{gg} = 2.0$.
 In the figure, we also indicate excluded parameter region which violate perturbative condition 
 $y^{U,D} < 4 \pi$ and $\lambda_\Phi < 4 \pi$ derived from Eq.~(\ref{eq:yF}) and (\ref{eq:lamS}) respectively.
 Thus   a sizable production cross section can be obtained in perturbative parameter region.

\begin{figure}[tbh] 
\begin{center}
\includegraphics[width=60mm]{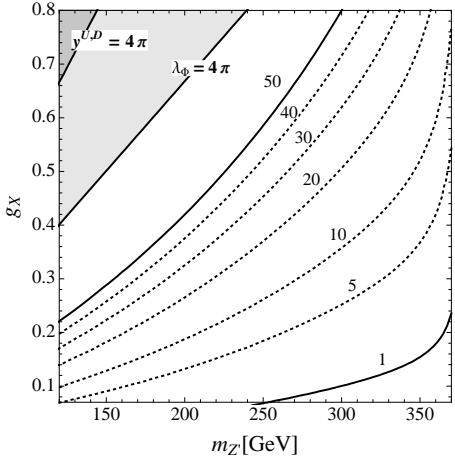} \hspace{5mm}
\caption{ The total decay width of $\phi$ in unit of GeV with same parameter setting as Fig.~\ref{fig:XS}. }
\label{fig:width}
\end{center}
\end{figure}

\begin{figure}[tbh] 
\begin{center}
\includegraphics[width=60mm]{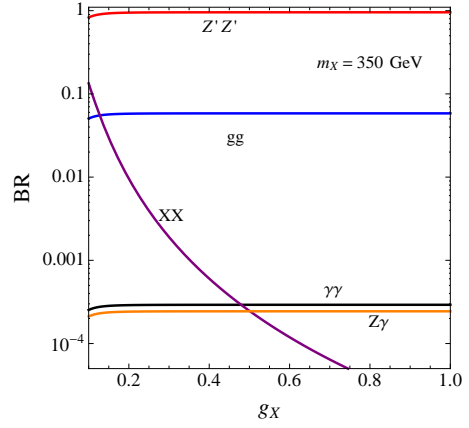} \hspace{5mm}
\caption{Branching fraction for decay of $\phi$. }
\label{fig:BR}
\end{center}
\end{figure}

\begin{figure}[tbh] 
\begin{center}
\includegraphics[width=60mm]{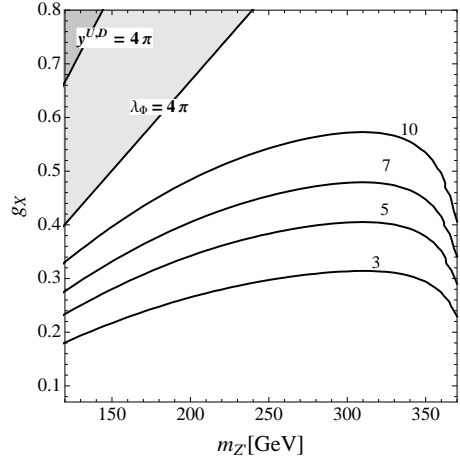} \hspace{5mm}
\caption{ The $\sigma (gg \to \phi) BR(\phi \to \gamma \gamma)$ in unit of fb with same parameter setting 
as Fig.~\ref{fig:XS}.}
\label{fig:XSBR}
\end{center}
\end{figure}

The partial decay widths for $\phi \to gg$ mode is derived by 
\begin{equation}
\Gamma_{\phi \rightarrow gg}  = \frac{ \alpha_s^2 m_\phi^3 }{32 \pi^3}  \left|\sum_{F=U,D}\frac{ (a+b) g_X}{2 m_{Z'}}    A_{1/2}(\tau_F) \right|^2.
  \end{equation}  
Similarly the partial decay width for $\phi \to \gamma \gamma$ is given via dark fermion loops such that
\begin{equation}
\Gamma_{\phi \to \gamma\gamma} = \frac{\alpha^2 m^3_\phi}{256 \pi^3} \left| \sum_F N_c^F \frac{(a+b)g_X Q^2_{F}}{m_{Z'}} A_{1/2}(\tau_F) \right|^2 \,,
  \end{equation}  
  where $Q_F$ and $N_c^F$ are electric charge and number of color of an exotic fermion $F$. 
The partial decay width for $\phi \to Z\gamma$ is also formulated by
\begin{align}
 \Gamma_{\phi \to Z\gamma} =& \frac{ m^3_\phi }{32\pi} \left| A_{Z\gamma}\right|^2  
\left(1 - \frac{m_Z^2}{m_\phi^2}\right)^3 \,, \\
 A_{Z\gamma} =& \frac{2\sqrt{2} \alpha s_W g_X}{\pi c_W} \nonumber \\
& \times \sum_{F} \frac{N_c^F (a+b) Q_{F}^2}{m_{Z'}} [I_1 ( \tau_{F}, \lambda_{F}) -I_2(\tau_{F}, \lambda_{F}) ]\,, \nonumber
\end{align}
where $\lambda_{F}= 4 m^2_{F}/m^2_Z$ and the loop integrals are given as~\cite{Gunion:1989we}:
 \begin{align}
 I_1(x,y) =& \frac{xy}{2(x-y)} + \frac{x^2 y^2}{2(x-y)^2} [ f(x)^2 - f(y)^2 ] \nonumber \\
& + \frac{x^2 b}{(x-y)^2} [g(x)-g(y)]\,,\nonumber \\
 I_2(x,y) =& - \frac{x y}{2(x-y)} [ f(x)^2 - f(y)^2 ] \,, \nonumber \\
 g(t) =& \sqrt{t -1} \sin^{-1}(1/\sqrt{t})\,. 
 \end{align}
On the other hand, the decay widths of $\phi$ into $Z'Z'$, $X^*X$ and $\bar F F$ modes are given 
at tree level as 
\begin{align}
\Gamma_{\phi \to Z' Z'} =& \frac{(a+b)^2 g_X^2 m_{Z'}^2}{32 \pi m_\phi} \nonumber \\
& \times \frac{m_\phi^4 - 4 m_\phi^2 m_{Z'}^2 + 12 m_{Z'}^4}{m_{Z'}^4} \sqrt{1 - \frac{4 m_{Z'}^2}{m_\phi^2}}\,, \\
\Gamma_{\phi \to X^* X} =& \frac{\lambda_{X\Phi}^2 m_{Z'}^2}{16 \pi (a+b)^2 g_X^2 m_\phi} 
\sqrt{1 - \frac{4 m_X^2}{m_\phi^2}}, \\
\Gamma_{\phi \to \bar F F} =& \frac{g_X^2 M_F^2}{4 \pi m_{Z'}^2} m_\phi \sqrt{1 - \frac{4 M_F^2}{m_{Z'}^2}}.
\end{align}
Fig.~\ref{fig:width} shows the total decay width of $\phi$ in the $m_{Z'} - g_X$ plane
 where the same parameter set as in Fig.~\ref{fig:XS} is used.
The branching fractions of $\phi$ decay can be obtained by partial decay widths, which is shown as 
a function of $g_X$ in Fig.~\ref{fig:BR} for $m_{Z'} = 300$ GeV with the above parameter setting.
Finally Fig.~\ref{fig:XSBR} shows contours of $\sigma( gg \to \phi) BR(\phi \to \gamma \gamma)$ 
in the $m_{Z'}-g_X$ plane.    We therefore find that $3-10$ fb cross section for diphoton mode can 
be obtained in the region of $g_X \simeq 0.2 - 0.5$ and $m_{Z'} < m_S/2$, simultaneously with a  
rather large decay width of $\phi$: $\Gamma_{\rm tot} (\phi) \approx 5-40$ GeV.

\subsection{Dark Matter Phenomenology}
The DM candidates of our model are $X$ and $N$.
We assume that the Higgs portal coupling $\lambda_{HX}=0$ for simplicity, since this case is
studied in great detail~\cite{Cline:2013gha}.  
We also assume that the Yukawa couplings involving the DM $X$ and SM fermions in Eq. (4) are
small enough so that their contribution to thermal relic calculation is negligible.  
Then the dominant annihilation processes of DM in our model are $XX^*(N \bar N)  \to Z' Z'$ assuming $m_{X,N} > m_{Z'}$.  We have included the $t(u)-$channel processes mediated 
by virtual $F$ exchange as well as the $s-$channel process mediated by $\phi$ exchange.   
Note that the $Z'$-exchanging processes are suppressed since interactions between $Z'$ 
and SM particles are small due to the small $Z-Z'$ mixing we assume.

\begin{figure}[tbh] 
\begin{center}
\includegraphics[width=60mm]{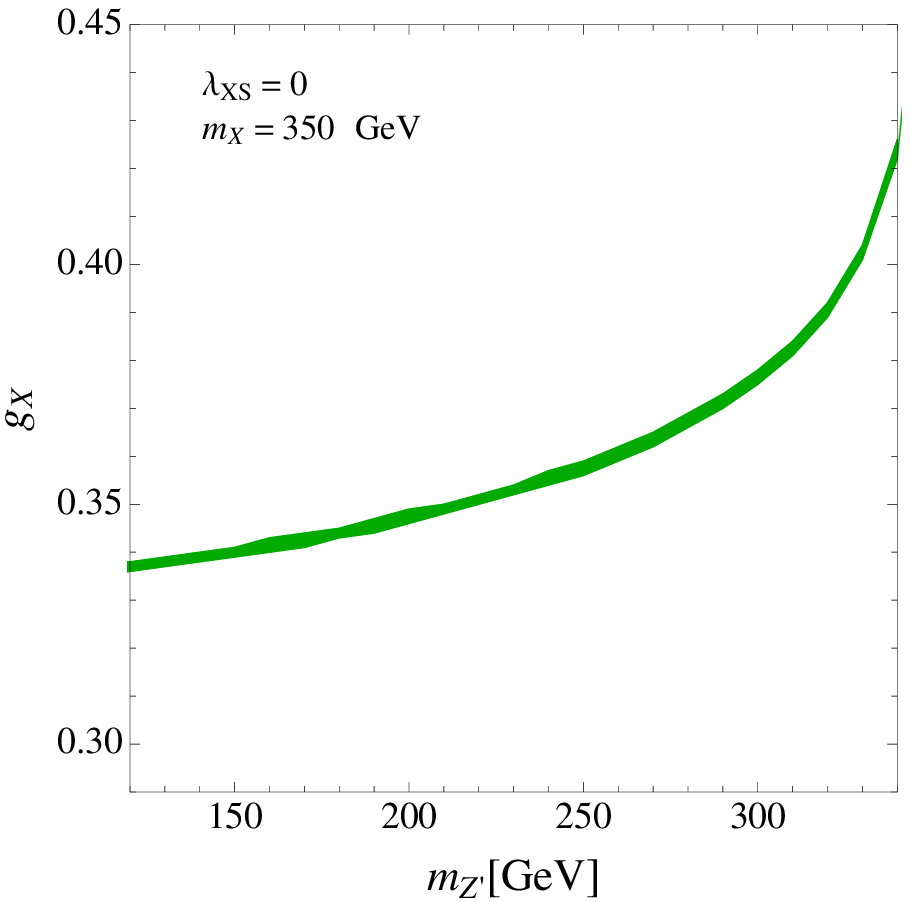} 
\includegraphics[width=60mm]{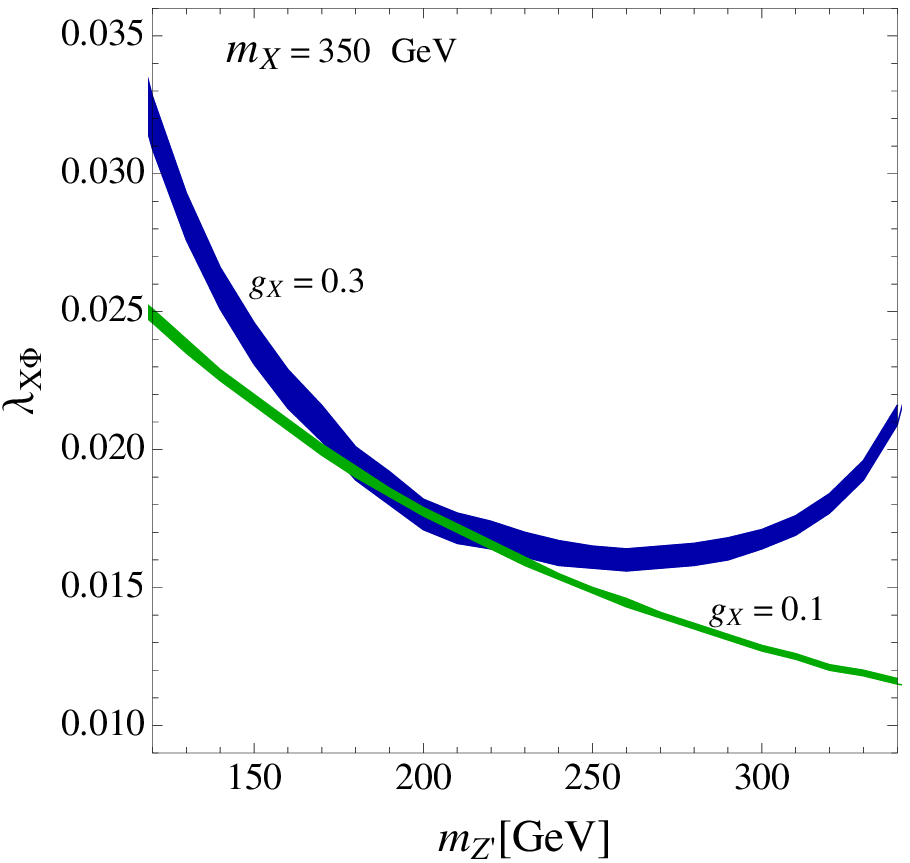} 
\caption{The colored region in upper(lower) plot indicate parameter space in $m_{Z'}-g_X (m_{Z'}-\lambda_{X\Phi})$ plane which explain 
observed relic density of $X$ where other parameters are indicated in the figures.}
\label{fig:RD}
\end{center}
\end{figure}

The thermal relic density is numerically estimated with { \tt micrOMEGAs 4.1.5 }
~\cite{Belanger:2014vza} to solve the Boltzmann equation by implementing relevant 
interactions relevant for the DM pair annihilation processes.
In calculating the relic density we assume $a \simeq b \simeq 1$ (but $a\neq b$).
We find that the DM relic density is given dominantly  by scalar DM $X$ in the parameter 
region where one can  explain the 750 GeV  diphoton excess. It turns out that the relic density 
of $N$ is small  due to large Yukawa coupling $y^N$ which makes the amplitude for the 
$\bar N N \to \phi \to Z' Z'$ process large.
Thus the thermal relic density of scalar DM $X$ is calculated with fixed parameter set of 
$\{M_{U,D}, M_{E,N}, m_X \} = \{800 \, {\rm GeV}, 400 \, {\rm GeV}, 350 \, {\rm GeV} \}$ 
and by taking $\{g_X, \lambda_{X \Phi }, m_{Z'} \}$ as free parameters.
We then search for the parameter region which give the right thermal relic density, i.e. 
$\Omega h^2 = 0.1199 \pm 0.0027$ as reported by Planck Collaboration~\cite{Ade:2013zuv}.
The upper figure in Fig.~\ref{fig:RD} shows the parameter region in the $( m_{Z'} , g_X )$ plane providing the observed relic density for $\lambda_{X \Phi} =0$.
On the other hand, the lower figure in Fig.~\ref{fig:RD} shows the corresponding parameter 
region in the $( m_{Z'} , \lambda_{X \Phi }) $ plane for $g_X =$ 0.1 and 0.3.
We find that interference between $t(u)-$ channel processes and $\phi$ exchanging $s-$channel process makes $\lambda_{X \Phi}$ dependence of the relic density nontrivial. 
For smaller $\lambda_{X\Phi}$ and $g_X$, small amount of Higgs portal coupling  
$\lambda_{HX}$ can help us to achieve the correct thermal relic density.

 In this model, DM-nucleon scattering occurs through $h$, $\phi$ and $Z^\prime$ exchanges. 
The amplitude for $Z^\prime$ exchange will be small since it involves $Z-Z^\prime$ 
mixing which can be sufficiently small. 
Also the Higgs contribution can be made small enough if we take a small $\lambda_{HX}$. 
For $\phi$ exchange, we have contribution to DM-nucleon scattering amplitude from 
$\phi$-gluon-gluon coupling in Eq.~(\ref{eq:LggS}) and $\phi-X-X$ coupling 
even if we suppress $\phi-h$ mixing.   The relevant effective coupling is given by
\begin{align}
{\cal L}_{XXGG} &= \frac{\alpha_S}{4 \pi} \left( \sum_{F=U,D} \frac{ \lambda_{X\Phi} }{m_{\phi}^2} A_{1/2}(\tau_F)  \right) X^\dagger X G^{a\mu \nu}G^a_{\mu \nu} \nonumber \\
& \equiv \frac{\alpha_S}{4 \pi} C_g X^\dagger X G^{a\mu \nu}G^a_{\mu \nu}.
\end{align}
Then the spin-independent DM-nucleon scattering cross section is obtained 
as~\cite{Giacchino:2015hvk} 
\begin{align}
\sigma_{\rm SI} &= \frac{m_N^2}{\pi (m_X + m_N)^2} f_N^2 \\
\frac{f_N}{m_N} &= - \frac{2}{9} C_g f^{(N)}_{T_G}
\end{align}
where $m_N$ is the nucleon mass and $f^{(N)}_{T_G}$ is the mass fraction of gluonic operators 
in the nucleon mass.  For the numerical values for these parameters, we adopt values in 
Ref.~\cite{Hisano:2015bma}. 
We find that DM-nucleon scattering cross section is small as $\sigma_{\rm SI} \lesssim 10^{-48} {\text cm}^2$ for the $\lambda_{X\Phi}$ providing the observed relic density in Fig.~\ref{fig:RD}.
Therefore it is difficult to observe the DM-nucleon scattering in direct detection experiment. 
%
%
%
\subsection{Muon $(g-2)_\mu$}

It is interesting to note that this model can also solve the muon $(g-2)_\mu$ 
through the dark muon and dark matter loop.   For $m_X = 350$ GeV and $m_{E_i} = 400$ GeV, 
we can account for the deficit in the $a_\mu =  8 \times 10^{-10} $ if $y^{E_i \mu} \sim 2-3$
assuming the universal $y^{E_i \mu}$ and $m_{E_i}$.  If we assume flavor conserving Yukawa,
$y \sim 5$ is needed. For such a large Yukawa coupling, however, we have large cross section for DM 
annihilation into lepton pair through the $t$-channel exchange  of $E_i$. 
Therefore when the muon $(g-2)_\mu$ is explained by the dark leptons within our model, the thermal relic
density of X is too small and we need another component of DM. Therefore we don't consider this possibility
any more in this letter.
 

\subsection{Stability of the potential}

Here we briefly discuss the stability of the scalar potential.
The one-loop beta functions of the Yukawa coupling $y^F$ and $\lambda_\phi$ are given by~\cite{Son:2015vfl}
\begin{align}
\label{eq:run1}
\beta_{y^F} &= y^F \left[ 3(2 N_c^F +1) (y^F)^2 - \frac{18}{5} Q_F^2 g_1^2 - 8 g_3^2 \right], \\
\label{eq:run2}
\beta_{\lambda_\Phi} &= 8 \lambda_\Phi \sum_F N_c^F (y^{F})^2 + 18 \lambda_\Phi^2 
- 8 \sum_F N_c^F (y^F)^4
\end{align}
where $g_{1(3)}$ are gauge couplings for SU(1)$_Y$(SU(3)) and the $\overline{\text MS}$ 
scheme is applied.   As a rough estimation, we ignore the running of gauge 
couplings in the energy range of $O(1)$ TeV to $O(10)$ TeV since the moderate running of gauge couplings in the RHS of Eq.~(\ref{eq:run1}) does not make significant changes for the running behavior of $y^F$ and $\lambda_\Phi$. 
\begin{figure}[tbh] 
\begin{center}
\includegraphics[width=60mm]{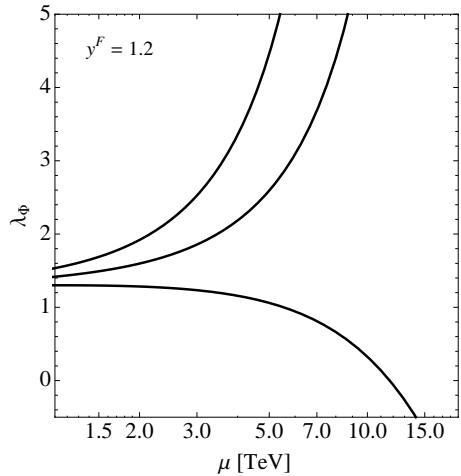} 
\caption{The running of $\lambda_\Phi$ according to Eq.~(\ref{eq:run1}) and (\ref{eq:run2}) where we adopted $y^F = 1.2$ and $\lambda_\Phi = \{1.3, 1.4,1.5 \}$ at $\mu = 1$ TeV as reference points.}
\label{fig:running}
\end{center}
\end{figure}
In Fig.~\ref{fig:running}, we show the renormalization group running of $\lambda_\Phi$ where we took $\lambda_\Phi = \{1.3,  1.4, 1.5 \}$ as reference points at $\mu = 1$ TeV 
 and assumed universal Yukawa couplings $y^F = 1.2$ at the same $\mu$ for simplicity.
We thus find that $\lambda_\Phi$ cannot be too small or too large to stabilize the potential.
Also relative magnitude between $y^F$ and $\lambda_\Phi$ changes the running property significantly, which can be tuned by changing U(1)$_X$ 
 charge of $\Phi$, $a+b$, 
according to Eq.~(\ref{eq:yF}) and (\ref{eq:lamS}).
By tuning the parameters, the stability of the potential can be achieved up to $\sim$ 10 TeV.
The complete analysis is beyond the scope of this letter and we left it as future work.

\subsection{Future Tests of This Model}

The model presented in this letter can be tested at the upcoming LHC experiments by searching 
for a pair of dark photons around $m_{Z^{'} Z^{'}} \sim 750$ GeV in the following channels:
\begin{eqnarray*}
p p & \rightarrow & \phi \rightarrow Z^{'} Z^{'} 
\\
Z^{'} Z^{'}  & \rightarrow & 4j \ , 2j + ll \ , 2j + \ET, 4 l \ , 2 l + \ET,
\end{eqnarray*}
where $\ET$ is from $\nu\bar{\nu}$ pair.  Note that the total decay width of dark photon $Z^{'}$ should
be very narrow, $\Gamma_{\rm tot} (Z^\prime) /m_{Z^\prime} \lesssim 10^{-4}$.
If the current ATLAS result on $\Gamma_\phi \sim 45$ GeV is confirmed in the future, our model predicts
that the main decay channel of dark Higgs $\phi$ should be a pair of dark photon, with a large cross section,
$\sigma (\phi \rightarrow Z^{'} Z^{'}  ) \approx O(5-40)$ pb (see Fig.~1) at the LHC@$\sqrt{s} = 13$ TeV.
Therefore a dedicated search for dark photon pair could confirm or exclude our model.

Our model also opens widely a new window for DM model building, especially the Higgs portal DM.
By assuming that the dark sector matter fields carry nonzero SM charges, the collider signatures become
richer and also the Higgs signal strength can be different from the usual Higgs portal DM models 
in the presence  of the mixing between the dark Higgs and the SM Higgs bosons.   Our model can satisfy 
all the constraints from (in)direct search bounds as well as DM searches at colliders.

\section{Conclusion}

In this letter, we proposed a new dark matter model with 3 generations of dark fermions 
that are chiral under new dark U(1)$_X$ gauge symmetry.   Both dark photon and the dark fermions get
their masses entirely from spontaneous breaking of dark U(1)$_X$ gauge symmetry from the nonzero 
VEV of $\Phi$, and dark Higgs boson $\phi$ appears as a result. 
Then the diphoton excess at 750 GeV is identified as the dark Higgs boson from U(1)$_X$ 
symmetry breaking.  The main decay mode of $\phi$ is a pair of  dark photon ($\phi \rightarrow Z^{'} Z^{'}$) 
and could be probed at the LHC by searching for $4j, 2j + ll, 2j + \ET, 4 l, 2l + \ET$.
It is remained to be seen if the 750 GeV diphoton excess survives in the future data accumulation.
If it does, the model presented in this letter would be an interesting possibility without conflict with the 
known experimental constraints even for large decay width of $\phi$.  In particular the production and 
the decay  of the dark Higgs boson $\phi$ involves dark fermions in the triangle loops, 
opening a new window to the dark  sector.
\begin{acknowledgments}
We are grateful to Jack Kai-Feng Chen, Sung Won Lee and Hwidong Yoo for discussions 
on the experimental status on the dark photon searches. 
 We also thank the anonymous referee for valuable suggestions. 
This work is supported in part by National Research Foundation of Korea (NRF) Research Grant NRF-2015R1A2A1A05001869, and by SRC program of NRF Grant No. 20120001176 funded by MEST through Korea Neutrino Research Center at Seoul National University (PK).
\end{acknowledgments}

\bibliography{basename of .bib file}

\end{document}